\documentclass[sigconf,nonacm=true,balance=true]{acmart}

\AtBeginDocument{%
  \providecommand\BibTeX{{%
    \normalfont B\kern-0.5em{\scshape i\kern-0.25em b}\kern-0.8em\TeX}}}

\setcopyright{none}

\usepackage{listings}
\usepackage{color}
\definecolor{light-gray}{gray}{0.95}
\definecolor{codegray}{rgb}{0.5,0.5,0.5}
\usepackage{enumitem}
\usepackage{hyperref}
\hypersetup{
pdftitle={From Proprietary to High-Level Trigger-Action Programming Rules: A Natural Language Processing Approach},
pdfsubject={Trigger-Action Programming Rules},
pdfauthor={Ekene Attoh and Beat Signer},
pdfkeywords={Internet of Things, Cross-Platform IoT, Natural Language Processing, End-User Authoring, Semantic Interoperability}
}

\usepackage{balance} 

\newcommand{\dquote}[1]{``#1''}

\newcommand{\code}[1]{\texttt{#1}}
\newcommand{\edquote}[1]{\emph{``#1''}}

\lstdefinestyle{mystyle}{
  backgroundcolor=\color{backcolour}, commentstyle=\color{codegreen},
  keywordstyle=\color{magenta},
  numberstyle=\tiny\color{codegray},
  stringstyle=\color{codepurple},
  basicstyle=\ttfamily\footnotesize,
  breakatwhitespace=false,         
  breaklines=true,                 
  captionpos=b,                    
  keepspaces=true,                 
  numbers=left,                    
  numbersep=5pt,                  
  showspaces=false,                
  showstringspaces=false,
  showtabs=false,                  
  tabsize=2
}

\AtBeginDocument{%
  \providecommand\BibTeX{{%
    Bib\TeX}}}

\begin{document}

\title[From Proprietary to High-Level Trigger-Action Programming Rules: An NLP Approach]{From Proprietary to High-Level Trigger-Action\\ Programming Rules: A Natural Language Processing Approach}

\author{Ekene Attoh}
\affiliation{%
  \institution{Web \& Information Systems Engineering Lab}
  \institution{Vrije Universiteit Brussel}
  \streetaddress{Pleinlaan 2}
  \city{Brussels}
  \postcode{1050}
  \country{Belgium}
}
\email{eattoh@vub.be}

\author{Beat Signer}
\affiliation{%
  \institution{Web \& Information Systems Engineering Lab}
  \institution{Vrije Universiteit Brussel}
  \streetaddress{Pleinlaan 2}
  \city{Brussels}
  \postcode{1050}
  \country{Belgium}
}
\email{bsigner@vub.be}


\begin{abstract}
  With the rise of popular task automation or IoT~platforms such as \emph{If~This~Then~That~(IFTTT)}, users can define rules to enable interactions between smart devices in their environment and thereby improve their daily lives. However, the rules authored via these platforms are usually tied to the platforms and sometimes even to the specific devices for which they have been defined. Therefore, when a user wishes to move to a different environment controlled by a different platform and/or devices, they need to recreate their rules for the new environment. The rise in the number of smart devices further adds to the complexity of rule authoring since users will have to navigate an ever-changing landscape of IoT devices. In order to address this problem, we need human-computer interaction that works across the boundaries of specific IoT~platforms and devices. A step towards this human-computer interaction across platforms and devices is the introduction of a high-level semantic model for end-user IoT~development, enabling users to create rules at a higher level of abstraction. However, many users who already got used to the rule representation in their favourite tool might be unwilling to learn and adapt to a new representation. We present a method for translating proprietary rules to a high-level semantic model by using natural language processing techniques. Our translation enables users to work with their familiar rule representation language and tool, and at the same time apply their rules across different IoT~platforms and devices.
\end{abstract}




\maketitle

\section{Introduction and Related Work}
\label{sect:int_rel_work}

The interoperability issue between devices of different brands in the domain of the Internet of Things~(IoT) is still ever present, a main reason being the unwillingness of major device manufacturers to provide a means for their devices to interoperate with their competitors' devices~\cite{longo2022towards}. 
Noura~et~al.~\cite{noura2019interoperability} studied the interoperability gaps present in state-of-the-art IoT~solutions. They defined different categories to classify IoT~interoperability issues and found that most IoT~solutions do not support \emph{cross-platform} and \emph{cross-domain} interoperability. If supported, these categories can enable IoT~users to exploit different IoT~services independently of the platform (e.g.~Apple or Samsung) or domain (e.g.~health or mobility). End-user development has been suggested as a means to give users control over IoT~solutions. According to Barricelli~et~al.~\cite{barricelli2019end}, end-user~development enables users to develop and adapt systems at a level of complexity that is adequate to their background and skills.

Various tools have been proposed to support end-user IoT~development~\cite{desolda2016end,Coutaz16,ospan2018context}.
Markopoulos~et~al.~\cite{markopoulos2017end} noted that the most common programmatic end-user control of IoT~applications is through specifying rules. Previous studies~\cite{cabitza2017rule,ur2016trigger} have further shown that a rule-based approach is easily understandable and enables end users to create their own programs. Li~et~al.~\cite{li2017programming} identified that although the popular IoT~task automation platform IFTTT\footnote{\url{https://ifttt.com}} enables users to create rules across various devices and services, only a certain number of (partner) devices and services are supported.

Corno~et~al.~\cite{corno2021devices} reiterate this by stating that most end-user development platforms adopt a vendor-centric abstraction, thus requiring that every online service needs to be programmed in a specific way. They argue that this poses interoperability challenges since users need to know any technological details to execute the intended behaviours beforehand. This approach is inadequate in future IoT~environments like smart cities, as \emph{things} will not always be known a priori but might dynamically appear and disappear~\cite{corno2021devices}.

To address this interoperability issue, Li~et~al.~\cite{li2017programming} presented a solution allowing users to author rules for their IoT~devices by demonstrating interactions between smart devices using their mobile phones. The solution consists of an \emph{Android} application enabling users to create automation scripts. These scripts are composed by recording the actions users perform on the mobile application of their smart devices. The scripts can then be triggered to perform the actions which were demonstrated by the user, using a source from another application such as a notification from a motion sensor application. We note that since the solution was designed to work on the Android operating system, the created automations will only be usable on Android devices. This implies that a user would lose all their automations if they were, for example, going to switch to an iOS~device. Further, only IoT~devices whose manufacturers provide an Android application for remote control are supported.

Corno~et~al.~\cite{corno2019high} attempted to address the interoperability issue by introducing the \emph{EUPont} (End User Programming Ontology) high-level semantic model for end-user IoT~development. With EUPont, users no longer need to create rules for specific devices or services, but they can define abstract rules such that any device or service able to perform the required action can be used to execute those rules. For example, a specific IFTTT~rule like \edquote{If my smart sensor X detects that I am home and the outside temperature is less than 10 degrees, then turn on my smart heater H} would have to be recreated if a user switches to a smart sensor \edquote{Y} or smart heater \edquote{I}. With EUPont, the rule can be transformed to \edquote{If I am in an indoor place and the outside temperature is less than 10 degrees, then start heating the indoor place}. Using any platform understanding the EUPont ontology, a user's rules can be executed on any device that can detect the user's presence and also heat up their environment. A user study revealed that the EUPont representation allows end users to reduce errors and the time needed to compose their IoT~applications.

Another factor contributing to the problem of interoperability in~IoT is an issue present in end-user IoT development. As more technologies are supported by platforms such as IFTTT, the design space also grows and it becomes more difficult for users to discover rules and their related functionality~\cite{corno2020taprec}. This increased complexity will lead to the creation of multiple rules with similar functionality realised in different possible ways, thus making interoperability more difficult. The use of recommendations in end-user development tools has been proposed to address this issue, similar as for the general development of software artefacts~\cite{kruger2018extract,nguyen2016api}.

However, these opportunities for recommendations have not yet been consistently explored to support end-user development but rather focus on supporting professional developers~\cite{corno2020taprec}. Therefore, the \emph{TAPrec} end-user development platform supporting the composition of trigger-action rules based on dynamic recommendations has been introduced. At composition time, it suggests new rules to be used or actions, which are based on the rule's final purpose such as illuminating a place rather than details like device brands and manufacturers, for completing a rule.

Mattioli~et~al.~\cite{andrea2021} proposed a solution that suggests relevant triggers, operators and actions to a user during rule composition. The system provides both, step-by-step and full-rule recommendations and a user is either recommended components to complete their rule or the system suggests a complete rule. Jeong~et~al.~\cite{jeong2019big} introduced a framework to analyse the usage logs of devices in an IoT~context and make rule recommendations to users based on the analysis of their device usage patterns. Their solution is further able to make recommendations based on the usage pattern of other users who are in a similar context.

Huang~et~al.~\cite{huang2016instructablecrowd} proposed \emph{InstructableCrowd}, a crowd-sourcing system enabling users to create IF-THEN rules based on their needs. Based on a smartphone user interface, users can describe their problems---such as often being late for a meeting---to \emph{crowdworkers}, and the crowdworkers can then create rules addressing a user's needs and send them back to their phone. \emph{HeyTAP}~\cite{corno2020heytap} enables users to describe the desired behaviour of their smart devices to their system through conversations (text or voice) and get rule recommendations materialising their stated intentions.

Note that the \emph{situation} concept proposed by Trullemans~et~al.~\cite{trullemans2017context} and implemented in the Context Modelling Toolkit~(CMT) is another means to tackle the complexity of authoring rules with similar functionality. They proposed that the \emph{trigger} side of a rule can lead to the definition of a reusable \emph{situation} rather than just triggering an action. For instance, the rule \emph{\dquote{If my heart rate is higher than 100~bpm}} could lead to the situation \emph{\dquote{My heart rate is high}}. A situation can then further be used on the \emph{trigger} side of a new rule definition. This eliminates the need for another user to understand all the low-level details and they could simply use the situation \emph{\dquote{My heart rate is high}} in their own automations.

Although the presented related work proposed solutions to address IoT~interoperability issues, to the best of our knowledge they only propose solutions focusing on the creation of \emph{new} rules by users. Most research is focused on enabling users to create (new) rules in novel ways and using \emph{new} systems to bridge the interoperability gap rather than enabling users to retain their current tools and methods, while still being able to benefit from solutions that offer cross-platform interoperability~\cite{attoh2021middleware}. Based on our analysis of related work, we identified two major problems to be addressed:\vspace{0.2cm}

\noindent\textbf{Loss of Tooling Choice}:
As mentioned before, various solutions have been put forward to bridge the cross-platform gap~\cite{li2017programming,corno2019high}, but they also propose the use of new tools and languages. This means that users will need to learn to use new tools and rule descriptions in order to create rules that can be used across different platforms.\vspace{0.2cm}
    
\noindent\textbf{Rule Authoring Complexity}:
Related work further shows that due to the rise in the number of smart devices, the discoverability of rules and their related functionality becomes more complex~\cite{corno2020taprec}. Therefore, users do not only need to use new tools and rule description languages to benefit from cross-platform interoperability solutions, but they also have to navigate an ever-changing landscape of IoT~devices and services while authoring their rules. This additional complexity may not only create an entry barrier for new users, but also increase a user's time needed to create their desired automation. We analysed the IFTTT user recipes (rules) from the May~2017 dataset of Mi~et~al.~\cite{mi2017empirical} and found that out of the total 279'828 user recipes, there were 863~duplicate triggers (total number of triggers that were used more than once) and 502~duplicate actions (total number of actions that were used more than once). The identification of these triggers and actions, and the understanding of the underlying functionality is not a very complex task if they are relatively simple, but can easily become more difficult for more complicated or lesser-known triggers and actions. Dealing with that complexity in order to enjoy cross-platform intelligibility might be unmanageable for most users and in particular for non-expert users.

\begin{figure*}[htb]
  \centering
    \includegraphics[width=0.57\textwidth]{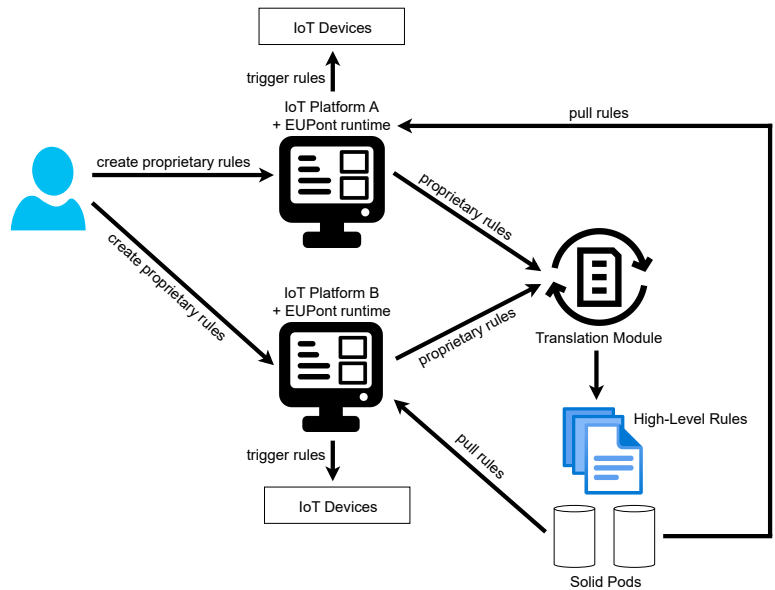}
    \caption{Proprietary IoT rule translation technique}
    \label{fig:interim-solution}
\end{figure*}

\section{Solution}
\label{sec:solution}
We propose a Natural Language Processing~(NLP) approach for automatically translating proprietary end-user rules to the \mbox{EUPont} high-level abstraction by Corno~et~al.~\cite{corno2019high}. This method provides end users with a \emph{Write Once, Run Anywhere} paradigm where a user can retain the authoring tool and language description of their choice but has the flexibility to use their rules across different platforms. A user simply has to write their rule as they would normally and have it translated to an equivalent EUPont representation. For instance, let us assume that a user has previously composed the IFTTT rule \edquote{If AC brand~X is turned off, then activate my camera brand~Y} for their smart home. They now find themselves on vacation in a smart environment which uses an air conditioner~(AC) of brand~Z and a camera of brand~C. With current solutions, this user needs to create a new rule \edquote{If AC brand~Z is turned off, then activate my camera brand~C} in order to have the same experience in their vacation environment as they would enjoy at their home. Instead, we propose a solution where a rule can be automatically translated to the EUPont generalisation \edquote{If device turned off, then connect to device}. Just as the JSON\footnote{\url{https://www.json.org}} format is generic such that most (modern) language compilers and interpreters are equipped with JSON parsers in order to work with JSON data, the intention behind the EUPont representation is similar such that IoT~platforms with the EUPont \edquote{runtime} might be able to work with the representation. Therefore, a proprietary rule such as an IFTTT rule has to be written only \emph{once} and can be translated to the EUPont representation in order to be used across different platforms. This means that an EUPont-powered platform can make it possible for any device which might be triggered off to be used as the \emph{trigger} of the rule. For the rule's action, a camera can be mapped to the high-level action \edquote{Connect to device}, which can then be triggered when the rule is executed. A user is therefore not limited to using devices of brand X or Y in order to take advantage of their already composed rule. 

An overview of our proposed IoT~rule translation approach is provided in Figure~\ref{fig:interim-solution}. A user is able to create (proprietary) IoT~rules, which can be seen as the \emph{Write Once} part, using any platform of their choice (e.g.~\code{IoT Platform~A} or \code{IoT~Platform~B}). These platforms have access to our translation approach described in the remaining part of this paper and the \code{Translation Module} then converts the created rules to the high-level EUPont representation.

As stated previously, a user would need to duplicate and further customise a rule authored to work on a specific platform in order to use that rule on a different platform, given that each platform stores its users' data locally. In order to address this issue, we also introduce the use of \code{Solid Pods} to store the automatically translated high-level rules. Solid aims to provide data independence as well as simple yet powerful data management mechanisms. Using Solid, applications no longer store their data themselves but request access to retrieve data from users' Pods~\cite{sambra2016solid}. The Solid integration thus enables a user to use these high-level rules on any IoT platform by granting individual platforms access to their Pod. This, in combination with the translation step, leads to the \emph{Run anywhere} part of our proposed solution. A user can write their rules using a proprietary platform such as IFTTT, these rules are then translated by our \code{Translation Module} and stored centrally on the user's Solid Pod. A platform can then pull the user's translated rules from the Solid Pods and with the help of the \code{\mbox{EUPont} runtime}, trigger the required actions on the corresponding IoT devices. This also puts users in control of their data, with their Solid Pods as the single access point and source of truth concerning their IoT rules.

Ultimately with our proposed translation solution, we aim to minimise or even eliminate a user's need to search for and/or understand the equivalent EUPont representation for their proprietary rules. In order to achieve this goal, we first propose that the user should be kept in the loop such that initially, they might manually select the best matching translation in situations where the one proposed by the system is inadequate. With this method, we intend that the best results are learned over time and then proposed to the user. Due to the popularity of the IFTTT platform, we have selected IFTTT rules as the first type of input to be addressed by our translation method. In future work, we intend to apply our approach to other IoT~platforms such as Home Assistant\footnote{\url{https://www.home-assistant.io}}.

As mentioned earlier, the May~2017 dataset of Mi~et~al.~\cite{mi2017empirical} contains 279'828 user recipes, implying that each of those recipes would need to be created for each new platform it is used on. With our proposed solution, these recipes can not only be written just once and then be automatically translated to run anywhere, but the author of a recipe can use their preferred authoring tool and keep the ownership of their data~(rules). In the remainder of this paper, we describe the translation approach depicted as \code{Translation Module} in Figure~\ref{fig:interim-solution}.

\subsection{Dataset}
\label{sub-sect:dataset}
Mi~et~al.~\cite{mi2017empirical} collected published IFTTT recipes (rules) from November 2016 until May 2017. For our automatic translation of rules, we decided to use the most recent May 2017 dataset containing a total of 279'828~recipes (rules). Note that not each of the collected recipes does necessarily contain unique triggers and actions. For example, the trigger \edquote{Any new photo by you} is used 9680~times in this dataset. Therefore, the 279'828 rules of the collected dataset consist of a total of~1017 different triggers and 616~different actions of which~154 triggers and~114 actions appeared only \emph{once}. For our proof-of-concept implementation and evaluation, we ran our solution on all different triggers and actions.

\subsection{Data Preparation}
Before applying our translation technique, we performed some cleaning of the data in order to remove any present anomalies and to prepare it to be used for the translation steps. For the recipes (rules) present in Mi~et~al.'s~\cite{mi2017empirical} dataset, we noticed that some of the triggers and actions contained a forward slash character. We thus removed that character from the triggers and actions and further separated the triggers and actions into two different lists. In the following, we refer to these lists as the \emph{IFTTT dataset}. We also transformed the ontology proposed by Corno~et~al.~\cite{corno2019high} from XML to JSON in order to extract its high-level triggers and actions. We removed the \emph{Trigger} and \emph{Action} suffixes from the triggers and actions as they were redundant and separated the high-level trigger and action names into two different lists. In the following, we refer to these lists as the \emph{EUPont dataset}.

\subsection{Translation Technique}
\label{sub-sect:technique}
Our aim in translation was to take a rule written by a user in a proprietary format (e.g.~IFTTT) and return a generalisation of that rule in the high-level EUPont~format by Corno~et~al.~\cite{corno2019high} that is as accurate as possible. This would enable the user to maintain their use of the IFTTT platform without having to learn a new rule description language. In order to perform this automatic translation from proprietary rules to high-level rules, we apply some natural language processing~(NLP) techniques. According to Quarashi~et~al.~\cite{qurashi2020document}, measuring text similarity is an important part of NLP~applications, such as information retrieval, machine translation and text summarisation. In order to perform the translation, we applied different \emph{document similarity} algorithms to both, the IFTTT and EUPont datasets, using the algorithm shown in Listing~\ref{pseudocode}.

\begin{lstlisting}[caption={Translation algorithm pseudocode}, captionpos=b, backgroundcolor=\color{light-gray}, label={pseudocode},basicstyle=\scriptsize]
for each trigger x in EUPont dataset:
  for each trigger y in IFTTT dataset:
    run document_similarity(x,y)
return x, y, Similarity(x,y) 
order by similarity descending

for each action a in EUPont dataset:
  for each action b in IFTTT dataset:
    run document_similarity(a,b)
return a, b, Similarity(a,b) 
order by similarity descending
\end{lstlisting}

We used the algorithm with three implementations (spaCy, AlleNLP and combined similarity) of the \code{document\_similarity(x,y)} function in Listing~\ref{pseudocode} and compared the results as described.

\noindent\textbf{spaCy Similarity}:
\emph{spaCy}\footnote{\url{https://spacy.io/usage/spacy-101}} is a free open source Python library for advanced Natural Language Processing. It can be used to build information extraction, natural language understanding systems or even to pre-process text for deep learning. We use spaCy's \emph{similarity} feature to compare how similar a given IFTTT trigger and action are to the high-level triggers and actions presented in the ontology by Corno~et~al.~\cite{corno2019high}. We then return the IFTTT trigger or action name together with the computed similar EUPont trigger or action names, as well as the corresponding similarity level.

\begin{lstlisting}[caption={spaCy IFTTT trigger translation example}, captionpos=b, backgroundcolor=\color{light-gray}, label={spaCy_result}, basicstyle=\scriptsize]
[
  {
    "Every Time": {
      "ifttt_name": "Any event starts",
      "similarity": 0.7474772725000891 }
  },
  {
    "Every Day": {
      "ifttt_name": "Any event starts",
      "similarity": 0.7034427432691928 }
  },
  {
    "Every Week": {
      "ifttt_name": "Any event starts",
      "similarity": 0.6832991463006063 }
  },
  {
    "Every Year": {
      "ifttt_name": "Any event starts",
      "similarity": 0.6819517479034135 }
  }
]
\end{lstlisting}

An example of a result is shown in Listing~\ref{spaCy_result}, where the first entry has the EUPont trigger \edquote{Every Time} returned by the spaCy~approach for the IFTTT trigger name \edquote{Any event starts}. Since not all translations returned by the algorithm are relevant, we defined a \emph{threshold value} which is used to filter out results whose similarity falls below that value. While the threshold is customisable, based on our initial analysis we set its value to~0.55.\vspace{0.1cm}

\noindent\textbf{AllenNLP Similarity}:
AllenNLP~\cite{Gardner2017AllenNLP} is an entire platform for solving NLP tasks and comes with a Python library. We applied the \emph{textual entailment} feature of AllenNLP which, for a pair of sentences, predicts whether the facts in the first sentence imply the facts in the second. We thus determine the textual entailment between each IFTTT trigger and action in the dataset, and the high-level EUPont~\cite{corno2019high} triggers and actions. The AllenNLP textual entailment algorithm returns \emph{entailment} (a measure of the similarity of both texts), \emph{contradiction} (a measure of the dissimilarity of both texts) and \emph{neutral} (a measure of the neutrality of both texts). We return the IFTTT triggers and actions together with the computed similar \mbox{EUPont} triggers and actions as well as their entailment, contradiction and neutral values as illustrated in Listing~\ref{allennlp_result}.
    
\noindent\textbf{Combined Similarity}:
Our preliminary analysis revealed that the spaCy approach returns more reliable results than the AllenNLP approach. In order to improve the translation results and reduce any noise, we defined a new approach where the AllenNLP algorithm is used to compare the similarity between the initial spaCy results and the EUPont triggers and actions. For example, for the IFTTT trigger \edquote{Any event starts}, we see that the first spaCy result returned is \edquote{Every Time} while the first AllenNLP result returned is \edquote{Taken}.

\begin{lstlisting}[caption={AllenNLP IFTTT trigger translation example}, captionpos=b, backgroundcolor=\color{light-gray}, label={allennlp_result},basicstyle=\scriptsize]
[
  [
    {
      "ifttt_name": "Any event starts",
        "eupont_hypothesis": "Taken",
        "allen_nlp_entailment": 92.16328263282776,
        "allen_nlp_contradiction": 3.0818356201052666,
        "allen_nlp_neutral": 4.75488156080246
    },
    {
      "ifttt_name": "Any event starts",
        "eupont_hypothesis": "Received",
        "allen_nlp_entailment": 91.5201187133789,
        "allen_nlp_contradiction": 3.172384202480316,
        "allen_nlp_neutral": 5.307500064373016
    },
    {
      "ifttt_name": "Any event starts",
        "eupont_hypothesis": "Temporal",
        "allen_nlp_entailment": 88.79325985908508,
        "allen_nlp_contradiction": 5.660351365804672,
        "allen_nlp_neutral": 5.546396225690842
    },
    {
      "ifttt_name": "Any event starts",
        "eupont_hypothesis": "Received From Diy",
        "allen_nlp_entailment": 86.96905374526978,
        "allen_nlp_contradiction": 2.1996214985847473,
        "allen_nlp_neutral": 10.831315815448761
    }
  ]
]
\end{lstlisting}
\vspace{0.2cm}

While the first result returned by the spaCy approach might be acceptable, the result returned by the AllenNLP approach is not. However, while conducting our preliminary analysis, we noted that there were more accurate matches further down in the list of results returned by the spaCy approach. We therefore ran the AllenNLP algorithm using the preliminary spaCy results and the EUPont triggers in order to further improve the translation.

In Listing~\ref{combined_result} we can see that the first result returned using this combined approach is \edquote{Started Activity}. This result is obtained by combining (averaging) its original spaCy similarity value~(61.39) with the similarity value obtained when using the AllenNLP algorithm~(85.80). The entry therefore has a combined similarity of~73.60. We are thus able to move this specific spaCy result---which had initially a low ranking compared to the initial \edquote{Every Time} top result---to the top. We perform this process for each result returned by the spaCy approach and therefore the results with a high combined similarity are most likely to be the most accurate since they have both, high spaCy and high AllenNLP similarities. An example of the result obtained using this combined approach is highlighted in Listing~\ref{combined_result}.

While there are several NLP libraries available for the Python programming language such as scikit-learn\footnote{\url{https://scikit-learn.org/stable/tutorial/text_analytics/working_with_text_data.html}} and PyTorch\footnote{\url{https://pytorchnlp.readthedocs.io/en/latest/}}, we opted for the spaCy\footnote{\url{https://spacy.io}} and AllenNLP libraries mainly due to their user friendliness. spaCy is also a popular choice in NLP~tasks given its speed and the ease with which it lets users build solutions. Similarly, AllenNLP enjoys popularity due to the speed and ease with which it lets a user build prototypes.

\begin{lstlisting}[caption={Combined IFTTT trigger translation example}, captionpos=b, backgroundcolor=\color{light-gray}, label={combined_result}, basicstyle=\scriptsize]
[
  {
    "ifttt_name": "Any event starts",
    "eupont_hypothesis": "Started Activity",
    "spacy_similarity": 61.39593233371522,
    "allen_nlp_entailment": 85.80678701400757,
    "allen_nlp_contradiction": 3.3948026597499847,
    "allen_nlp_neutral": 10.798408836126328,
    "combined_similarity": 73.6013596738614
  },
  {
    "ifttt_name": "Any event starts",
    "eupont_hypothesis": "Position Registration",
    "spacy_similarity": 57.52355435396419,
    "allen_nlp_entailment": 81.54605627059937,
    "allen_nlp_contradiction": 5.951366946101189,
    "allen_nlp_neutral": 12.50256896018982,
    "combined_similarity": 69.53480531228178
  },
  {
    "ifttt_name": "Any event starts",
    "eupont_hypothesis": "Device Turned On",
    "spacy_similarity": 56.46723924755545,
    "allen_nlp_entailment": 80.1530659198761,
    "allen_nlp_contradiction": 5.053842067718506,
    "allen_nlp_neutral": 14.793087542057037,
    "combined_similarity": 68.31015258371578
  },
  {
    "ifttt_name": "Any event starts",
    "eupont_hypothesis": "Time",
    "spacy_similarity": 67.30982538675227,
    "allen_nlp_entailment": 65.08164405822754,
    "allen_nlp_contradiction": 16.626055538654327,
    "allen_nlp_neutral": 18.29230487346649,
    "combined_similarity": 66.1957347224899
  }
]
\end{lstlisting}

\section{Results}
\label{sec:results}
We recorded and compared the results we obtained when applying each of the three different approaches described in Section~\ref{sub-sect:technique} on recipes of the Mi~et~al.~\cite{mi2017empirical} dataset. With our translation technique, we intended that the first result recommended by each approach should produce the most accurate high-level EUPont generalisation of the IFTTT triggers and actions. However, in situations where this is not the case, a user should at least be able to find an accurate high-level EUPont generalisation within the first five returned results. We decided to consider only the first five results by each approach to reduce the potential burden a user might face when looking for the best result. In our results, we mark an entry with \edquote{No result} if no suitable match has been returned as part of the first five results. For trigger or action names that we consider to be ambiguous---any trigger or action whose meaning could have multiple interpretations---we mark the entry and the resulting translations as \edquote{Ambiguous}. For instance, the trigger \edquote{Air quality changed} is marked as \edquote{Ambiguous} because the change in air quality could either be positive or negative. Therefore one approach might return \edquote{Air quality decreased} as its best result, while another approach might return \edquote{Air quality increased} as its best result. We further note that several possible acceptable results were returned for certain IFTTT triggers and actions by our approach in the different columns.

\begin{table*}[htb]
    \resizebox{\textwidth}{!}{%
    \centering
    \begin{tabular}{ p{0.02\linewidth}p{0.14\linewidth}p{0.16\linewidth}p{0.16\linewidth}p{0.09\linewidth}p{0.10\linewidth}p{0.16\linewidth}p{0.16\linewidth} }
        \hline
        \textbf{No.} & \textbf{IFTTT Name} &  \textbf{spaCy 1} & \textbf{spaCY 2} & \textbf{AllenNLP 1} & \textbf{AllenNLP 2} & \textbf{Combined 1} & \textbf{Combined 2} \\
        \hline
        1 & \ldots & \ldots & \ldots & \ldots & \ldots & \ldots & \ldots \\
        \hline
        2 & A C~\footnote{A C means Air Conditioning} turned off & Device Turned Off & Device Turned Off~(1) & Brightness Decreased & No result & Device Turned Off & Device Turned Off~(1) \\
        \hline
        23 & Action Button\newline Pressed & Tap Button Activity & Tap Button Activity~(1) & Taken & No result & Tap Button Activity & Tap Button Activity~(1) \\
        \hline
        34 & Air filter needs\newline cleaning & Air Purifier Enabled & No result & Moving & No result & Started Cleaning & Sensed Air Quality\newline Decreased~(3) \\
        \hline
        35 & Air pressure drops\newline below & Increased Air Pressure & Sensed Air Pressure\newline Decreased~(4) & Temporal & No result & Sensed Air Pressure\newline Decreased & Sensed Air Pressure\newline Decreased~(1) \\
        \hline
        36 & Air pressure rises\newline above & Increased Air Pressure & Increased Air\newline Pressure~(1) & Temporal & No result & Sensed Air Pressure\newline Increased & Sensed Air Pressure\newline Increased~(1) \\
        \hline
        37 & Air purifier is\newline turned on & Air Purifier Enabled & Device Turned On~(4) & Taken Audio & Device Turned On~(2) & Device Turned On & Device Turned On~(1) \\
        \hline
        38 & Air quality changed\newline (ambiguous) & Increased Air Quality & Ambiguous & Temporal & Ambiguous & Increased Air Quality & Ambiguous \\
        \hline
        48 & An animal has been\newline seen outside & Received Like & No result & Taken & Taken\newline Image~(2) & Taken & Taken Image~(2) \\
        \hline
        
    \end{tabular}}
    \caption{IFTTT trigger translation results}
    \label{tab:summary_triggers}
\end{table*}

\begin{table*}[htb]
    \resizebox{\textwidth}{!}{%
    \centering
    \begin{tabular}{ p{0.02\linewidth}p{0.14\linewidth}p{0.16\linewidth}p{0.16\linewidth}p{0.09\linewidth}p{0.10\linewidth}p{0.16\linewidth}p{0.16\linewidth} }
        \hline
        \textbf{No.} & \textbf{IFTTT Name} &  \textbf{spaCy 1} & \textbf{spaCY 2} & \textbf{AllenNLP 1} & \textbf{AllenNLP 2} & \textbf{Combined 1} & \textbf{Combined 2} \\
        \hline
        1 & \ldots & \ldots & \ldots & \ldots & \ldots & \ldots & \ldots \\
        \hline
        10 & Add a file & Share File & Save File~(2) & Information & No result & Share File & Save File~(2) \\
        \hline
        11 & Add a new site\newline (ambiguous) & Connect To Web\newline Service & Ambiguous & Start\newline Focusing & Ambiguous & Save Media\newline Information & Ambiguous \\
        \hline
        17 & Add a private\newline bookmark & Save Web Bookmark & Save Web\newline Bookmark~(1) & Information & No result & Get & Save Media\newline Information~(4) \\
        \hline
         31 & Add item to\newline Reading List & Add Calendar Item & Add Remind~(2) & Information & Save~(2) & Share Post & Add Remind~(3) \\
        \hline
        32 & Add item to your\newline feed & Add Calendar Item & No result & Start Cooking & No result & Get & Save Media\newline Information~(2) \\
        \hline
        33 & Add message & Send Message & No result & Information & No result & Add Reminder & No result \\
        \hline
        34 & Add mix to\newline favorites & Add Remind & No result & Get & No result & Add Remind & No result \\
        \hline
        35 & Add new contact\newline to list & Send To Display & Save Contact~(5) & Information & No result & Save Media\newline Information & Save Media\newline Information~(1) \\
        \hline
         
    \end{tabular}}
    \caption{IFTTT action translation results}
    \label{tab:summary_actions}
\vspace{-0.2cm}
\end{table*}

We split our findings into two different tables, with Table~\ref{tab:summary_triggers} showing the results for triggers and Table~\ref{tab:summary_actions} highlighting the results for actions. Each table contains the following columns:

\begin{itemize}[leftmargin=0.35cm]
    \item \textbf{No.}: Entry Number
    \item \textbf{IFTTT Name}: IFTTT trigger or action name
    \item \textbf{spaCy 1}: EUPont trigger or action name with the highest similarity value using the spaCy similarity algorithm.
    \item \textbf{spaCY 2}: EUPont trigger or action name most accurately representing the IFTTT trigger or action name using the spaCy similarity algorithm (position in the result list in brackets).
    \item \textbf{AllenNLP 1}: EUPont trigger or action name with the highest similarity value using the AllenNLP text entailment algorithm.
    \item \textbf{AllenNLP 2}: EUPont trigger or action name most accurately representing the IFTTT trigger or action name using the  AllenNLP text entailment algorithm (position in the result list in brackets).
    \item \textbf{Combined 1}: EUPont trigger or action name with the highest similarity value using the combined approach.
    \item \textbf{Combined 2}: EUPont trigger or action name most accurately representing the IFTTT trigger or action name using the combined approach (position in the result list in brackets).
\end{itemize}

For the results presented in this paper, we randomly selected 50~triggers and actions from the results we obtained from running our approaches on the dataset. The results from these 50~triggers and actions were then manually analysed (e.g.~to identify and label the most accurate EUPont triggers and actions) in order to populate the entries in Table~\ref{tab:summary_triggers} and Table~\ref{tab:summary_actions}. Note that only a few representative entries from these two tables are shown but the entire tables as well as the complete but non-annotated results of the presented approaches are available online~\cite{attoh_2023_10033916}.

\subsection{Analysis}
\label{sub_sect:analysis}
We sought to determine which out of the three approaches is performing best. We consider an approach to be performing better than another approach based on a combination of the following criteria:

\begin{table*}[htb]
    \resizebox{0.7\textwidth}{!}{%
    \centering
    \begin{tabular}{ p{0.18\linewidth}p{0.18\linewidth}p{0.18\linewidth}p{0.18\linewidth} }
        \hline
         \textbf{Approach} &\textbf{First Result} & \textbf{Top Five Result} & \textbf{No result} \\
        \hline
        spaCy & 13 & 9 & 16 \\
        \hline
        AllenNLP & 2 & 12 & 24 \\
        \hline
        Combined & 16 & 13 & 9 \\
        \hline
         
    \end{tabular}}
    \caption{Summary of trigger translation results }
    \label{tab:summary_results_triggers}
\end{table*}

\begin{table*}[htb]
    \resizebox{0.7\textwidth}{!}{%
    \centering
    \begin{tabular}{ p{0.18\linewidth}p{0.18\linewidth}p{0.18\linewidth}p{0.18\linewidth} }
        \hline
         \textbf{Approach} &\textbf{First Result} & \textbf{Top Five Result} & \textbf{No result} \\
        \hline
        spaCy & 10 & 8 & 22 \\
        \hline
        AllenNLP & 1 & 11 & 28 \\
        \hline
        Combined & 8 & 19 & 13 \\
        \hline
         
    \end{tabular}}
   \caption{Summary of action translation results}
    \label{tab:summary_results_actions}
\end{table*}

\begin{itemize}[leftmargin=0.35cm]
    \item It has more top results than the other approach
    \item It has more top~5 results than the other approach
    \item It has fewer cases where a translation could not be found in the top~5 results compared to the other approach
\end{itemize}

In our analysis, entries which were marked as \emph{ambiguous} (12 out of the 50 randomly selected triggers) were not considered. Therefore, we found that for the remaining 38~triggers, our combined approach returned the best EUPont match as the first result for 16 of those triggers as summarised in Table~\ref{tab:summary_results_triggers}. For 13 of the triggers, the best EUPont match was not the first result but could be found in the top 5~results. However, for 9 of the IFTTT triggers, a suitable EUPont match could not be found by our combined approach. Using the spaCy~approach, we found that the best EUPont match was returned as the first result for 13~triggers, while for 9~triggers, the best EUPont match was not the first result but could be found in the top~5 results. However, for 16~triggers, a suitable EUPont match could not be found. Similarly, using the AllenNLP approach we see that the best EUPont match was returned as the first result for 2~triggers. For 12 of the triggers, the best EUPont match was not the first result but could be found in the top~5 results, while for~24 of the triggers no suitable EUPont match could be found. 

For the actions, there were 10~entries marked as \emph{ambiguous}.
Table~\ref{tab:summary_results_actions} shows that when using our combined approach on the 40~considered actions, for 8 of the actions the best EUPont match was returned as the first result, while for 19 of the actions the best EUPont match was not the first result but could be found in the top~5 results and for 13 of the actions, no suitable EUPont match could be found. Using the spaCy approach, 10 of the actions had the best EUPont match returned as the first result, while for 8 of the actions the best EUPont match was not the first result but could be found in the top~5 results. However, for 22 of the actions, a suitable EUPont match could not be found. Using the AllenNLP approach, we see that only for 1 out of the actions the best EUPont match was returned as the first result, while for 11 of the actions the best EUPont match was not the first result but could be found in the top~5 results and for 28 of the actions, no suitable EUPont match could be found.

Based on these results, we can conclude that our combined approach is the best-performing approach using our test dataset for both triggers and actions. For triggers, our combined approach returns the highest number of top results, top~5 results and has the smallest number of cases where no result has been returned as part of the top~5. For the actions though, the spaCy approach returns more top results than our combined approach. However, spaCy returns significantly fewer top~5 results and has a larger number of cases where no result was returned as part of the top~5.

\section{Evaluation}
\subsection{EuPont Evaluation}
Corno~et~al~\cite{corno2019high} conducted a user study to evaluate the suitability and the understandability of the EUPont approach by end-users. The study was a controlled in-lab experiment that involved 30~participants, 15~of whom only had programming experience. It focused on the creation of IoT applications both with the current low-level representation of IFTTT and the high-level representation of \mbox{EUPont}. The study addressed the research questions \edquote{Does the EUPont representation help users create their IoT applications more effectively and efficiently compared to the low-level representation?} and \edquote{Which of the two representations is preferred by users, and which are the perceived advantages and disadvantages of the two solutions?}. In summary, the results of the study successfully demonstrated that the EUPont representation allowed end-users to reduce the errors and time needed to compose their IoT applications, and introduced numerous benefits in terms of understandability and ease of use. 

\subsection{User Evaluation}
\label{sec:user_evaluation}
Based on the findings of the user study conducted by Corno~et~al~\cite{corno2019high}, we consider the EUPont representation to be a suitable high-level representation of IoT rules for end users. To further evaluate our approach and gather some feedback for future work, we conducted a preliminary survey targeting a number of respondents who were already familiar with the use of IoT~automation solutions. With this survey, we aimed to investigate whether real IoT~users would find the results returned by any of our methods to be good high-level generalisations of the IFTTT triggers and actions included in the dataset described earlier in Section~\ref{sec:solution}. The research question we sought to answer with this preliminary survey was \edquote{Are the EUPont translations returned by our methods acceptable to end users?}
There were 11~survey respondents who started the survey but only two of them completed the survey. We thus base our analysis on the responses of these two respondents who were both male, aged between 20 and 39 years and have obtained a Master's degree. In our survey, the respondents were presented a series of IFTTT~triggers and actions with their corresponding translations based on the three approaches described in Section~\ref{sec:solution}. They then had to select which method they thought returned a good high-level generalisation of the IFTTT trigger or action and also specify to which degree they found that generalisation to be accurate. They could further select the \edquote{N/A} option in case they found that none of the methods returned a suitable generalisation. For example, given the IFTTT trigger \edquote{New photo upload on page}, users were presented the following four options:

\begin{itemize}[leftmargin=0.35cm]
    \item Method 1: \dquote{Shared Profile Update}
    \item Method 2: \dquote{No Result}
    \item Method 3: \dquote{Shared Post}
    \item \dquote{N/A}
\end{itemize}

In a follow-up question, they were asked \edquote{To which degree is your chosen method an accurate generalisation of the IFTTT trigger?} and had the choice of \edquote{Not at all accurate}, \edquote{Low accuracy}, \edquote{Accurate} and \edquote{Very accurate}. 

For situations where multiple methods returned the same value, the respondents were asked to pick any of those methods if they considered the value to be an accurate generalisation of the trigger or action. For example, for the IFTTT trigger \edquote{If new post from search \ldots}, method~1 and method~3 returned \edquote{If shared post \ldots} and therefore the respondents could choose any of those two methods if they found it a good high-level generalisation of the IFTTT trigger. We followed the same principle as described in Section~\ref{sub-sect:technique} in considering only the first five results of each method.

The triggers and actions selected for the survey were those we consider to be popular in the dataset; that is they were used 1000 or more times in the dataset described in Section~\ref{sec:solution}. Triggers and actions for which the three methods returned no suitable or ambiguous results were not selected. The survey thus comprised questions for 31~triggers and 33~actions. There were 20~out of the 31~triggers and 19~out of the 33~actions where more than one method returned the same value. We will refer to these cases as \emph{triggers with the same result} and \emph{actions with the same result} respectively. For 11~triggers and 14~actions none of the three methods returned the same result. We will refer to these cases as \emph{triggers with different results} and \emph{actions with different results} respectively.

For the 20~triggers with the same result, in 16~cases both respondents selected that result as a good high-level description, while for the other 4~cases they indicated that there was no suitable generalisation. In the case of the 19~actions with the same result, both respondents selected that result as a good high-level description in 14~cases, while for the other 5~cases they indicated that there was no suitable generalisation. For the 11~triggers with different results, there were five triggers where our combined method's result was not selected. For those five triggers, the result returned by the spaCy method was selected by both respondents; however, none of the other two methods returned a result for those five triggers. There were 5 of the 11~triggers with different results where the result of the combined method was selected by the respondents. Both respondents selected the result returned by the combined method in 3~out of those 5~cases, while for the other 2~cases, only one respondent selected the result returned by the combined method. The second respondent indicated that no suitable generalisation was found. For 3~out of these 4~triggers, the other methods did not return any suitable result. For one of the 11~triggers with different results, the result from our combined method was selected by one respondent while the result from the spaCy method was selected by the second respondent. Finally, there was one trigger where no method's result was selected even though our combined method was the only method that returned a result for that particular trigger.

In the case of the 14~actions with different results, there were 7~actions where our combined method's result was not selected. For those 7~actions, the result returned by the spaCy method was selected in 6~cases by both respondents and the result returned by the AllenNLP method was selected once by only one respondent. Our combined method only returned a result for 3~of those seven actions while AllenNLP returned a result for only one of the 7~actions. There were 4~of the 14~actions with different results where the result returned by the combined method was selected by the respondents. Both respondents selected the result returned by the combined method in 2~out of those 4~cases, while for the other 2~cases only one respondent selected the result returned by the combined method. The second respondent indicated that no suitable generalisation was found. For 3~out of these 4~actions, the other methods did not return any suitable result. For 2 of the 14~actions with different results, the result from our combined method was selected by one respondent while the result by the spaCy method was selected by the second respondent. Finally, there was one action where no method's result was selected even though our combined method was the only method that returned a result for that particular action. In summary, we can conclude that for 24~out of the 31~triggers and 22~out of the 33~actions, both respondents selected the result returned by one of our methods as a good high-level generalisation. For the remaining cases, both respondents did not select the result returned by one of our methods as a good high-level generalisation.

\section{Conclusions and Future Work}
We presented a \emph{Write Once Run Anywhere} paradigm for end-user authoring in IoT settings, helping users to maintain their preferred authoring tool as well as their preferred description language when defining IoT~rules that will work across different IoT~platforms. In order to achieve this, we employed the use of natural language processing techniques to automatically translate proprietary rules to high-level EUPont~rules. The findings of Corno~et~al~\cite{corno2019high} demonstrate that end users find the EUPont representation to improve the rule authoring process. We thus used two popular NLP algorithms in two different approaches and proposed a third novel approach by combining these two algorithms and carefully analysed the results obtained from all three approaches. From the results of our analysis in Section~\ref{sub_sect:analysis}, we see that all three methods return good high-level generalisations with the combined method performing better as described in~\ref{sub_sect:analysis} than the other two methods for the given dataset. We acknowledge that only a small set of users completed our survey, but from these preliminary results we see that real IoT~users also selected the results returned by our three approaches as good high-level generalisation for the triggers and actions that they were presented. We consider these results to be positive and for future work, we will investigate how to best consistently return the most accurate high-level generalisation for a user's rules, by either using one or a combination of the methods we have presented. We will also focus on improving the accuracy of the results such that more often the first result returned is the most accurate high-level generalisation and the number of cases where no result is returned is significantly minimised or even completely eliminated. Further, we plan to investigate how our solution can best be integrated into existing IoT~platforms as shown earlier in Figure~\ref{fig:interim-solution} in order to further evaluate the proposed NLP-based rule translation approach by end users.


\balance


\bibliographystyle{ACM-Reference-Format}
\bibliography{TR_Ekene_2023}

\end{document}